\documentclass[journal]{IEEEtran}
\normalsize

\usepackage[T1]{fontenc}

\usepackage{cite}

\ifCLASSINFOpdf
   \usepackage[pdftex]{graphicx}
   \DeclareGraphicsExtensions{.pdf,.jpeg,.png,.fig}
   \usepackage{everysel}
   \usepackage{keyval}
   \usepackage{ragged2e}
\else
   \usepackage[dvips]{graphicx}
   \DeclareGraphicsExtensions{.eps,.pdf,.jpeg,.png,.fig}
   \usepackage{caption}
   \usepackage[caption=false]{subfig}
   \usepackage{wrapfig}
\fi

\usepackage{epstopdf}

\usepackage{amsmath}
\usepackage{subfiles}
\usepackage{textcomp}
\usepackage[detect-weight=true, detect-family=true]{siunitx}
\usepackage{gensymb}
\usepackage{balance}
\usepackage{bm}
\usepackage{xr}
\usepackage[utf8]{inputenc}
\usepackage{array}
\usepackage{pbox}
\usepackage{cite}
\usepackage{subfigure}
\usepackage{booktabs}
\usepackage{xcolor}
\usepackage{wrapfig}
\usepackage{multirow}
\usepackage{dingbat}
\usepackage{amssymb} 
\usepackage{soul,color}
\usepackage[font=small,skip=0pt]{caption}

\usepackage{colortbl}

\usepackage{etoolbox}
\patchcmd{\thebibliography}
  {\settowidth}
  {\setlength{\itemsep}{0pt plus 0.1pt}\settowidth}
  {}{}
\apptocmd{\thebibliography}
  {\small}
  {}{
}

\begin{document}
%

\title{6G and Beyond: Joint Waveform and Wavefront Engineering To Realize Terahertz Communications}
\title{6G and Beyond: Wavefront Engineering to Realize Efficient Terahertz Band Communications in 6G and Beyond}
\title{Wavefront Engineering to Realize Efficient Terahertz Band Communications in 6G and Beyond}
\title{Wavefront Engineering: Realizing Efficient Terahertz Band Communications in 6G and Beyond}

\author{Arjun Singh, Vitaly Petrov, Hichem Guerboukha, Innem V.A.K. Reddy,\\
Edward W. Knightly, Daniel M. Mittleman, and Josep M. Jornet

\thanks{Arjun~Singh is with the SUNY Polytechnic Institute, USA. Vitaly~Petrov and Josep~M.~Jornet are with Northeastern University, USA. I.V.A.K. Reddy is with the University at Buffalo, USA. Hichem~Guerboukha and Daniel~M.~Mittleman are with Brown University, USA. Edward~W.~Knightly is with Rice University, USA.}
\thanks{The work has been supported in part by the National Science Foundation CNS-1955004, 1954780, 2148132, 2211616, 1955075, 1923782, 1824529, and 2148132, and the US Air Force Office of Scientific Research Award FA9550-23-1-0254 and FA9550-22-1-0412, and the Army Research Laboratory grant W911NF-19-2-0269. H.~Guerboukha also gratefully acknowledges support from Fonds de recherche du Quebec -- Nature et technologies.}
}


\maketitle

\begin{abstract}
Terahertz (THz) band communications is envisioned as a key technology for future wireless standards. Substantial progress has been made in this field, with advances in hardware design, channel models, and signal processing. High-rate backhaul links operating at sub-THz frequencies have been experimentally demonstrated. However, there are inherent challenges in making the next great leap for adopting the THz band in widespread communication systems, such as cellular access and wireless local area networks. Primarily, such systems have to be both: (i)~wideband, to maintain desired data rate and sensing resolution; and, more importantly, (ii)~operate in the massive near field of the high-gain devices required to overcome the propagation losses. In this article, it is first explained why the state-of-the-art techniques from lower frequencies, including millimeter-wave, cannot be simply repurposed to realize THz band communication systems. Then, a vision of wavefront engineering is presented to address these shortfalls. Further, it is illustrated how novel implementations of specific wavefronts, such as Bessel beams and Airy beams, offer attractive advantages in creating THz links over state-of-the-art far-field beamforming and near-field beamfocusing techniques. The paper ends by discussing novel problems and challenges in this new and exciting research area.
\end{abstract}


%
\IEEEpeerreviewmaketitle

\section{Introduction}
\label{sec:intro}

\IEEEPARstart{T}{he} terahertz (THz) band (0.3--3~THz) has been heavily investigated by the scientific community over the past decades for wireless sensing and communications. With developments relating to device technology, new waveform modulations, signal processing techniques, and robust digital back-ends, the so-called ``THz technology gap'' is slowly closing~\cite{akyildiz2022terahertz}. However, the open question that remains is how to design a practical THz band communication system that delivers on the promises of the next generations of wireless networks, which are tentatively marked as 1~terabit-per-second (Tbps) data rate, 0.1~ms latency, and ``$10^{-9}$'' reliability~\cite{akyildiz2022terahertz}.

Here, Shannon's capacity gives us major metrics to work with. More specifically, the channel capacity depends on: (i)~the available bandwidth, (ii)~the received signal-to-noise ratio (SNR), and (iii)~the spatial reuse factor. While the THz band provides massive available bandwidth for ultra-broadband modulations, the corresponding noise power, crippling path losses, and the fact that THz signals are easily obstructed by everyday objects can have devastating consequences on the SNR, leading to link breakage. The THz channel is also low rank, making many un-correlated multiple-input-multiple-output (MIMO) links unlikely~\cite{han2021hybrid}. 
These problems have thus spurred significant research efforts~\cite{di2019smart, akyildiz2022terahertz,zhou2022utilizing}.

To this end, in addition to the utilization of very high gain antenna arrays~\cite{balanis2016antenna}, intelligent reflecting surfaces (IRSs) are expected to support THz links~\cite{di2019smart}. The main goal with these devices is to utilize beamforming theory to increase the radiation gain, spatial multiplexing capability, and robustness of reconfigurable links including non-line-of-sight (NLoS) capability around obstacles. The very small wavelengths at THz frequencies result in such high-gain devices being relatively compact but having a massive near-field region. Thus, in most practical setups, the conventional far-field beamforming assumptions of the microwave, and to a large extent, the millimeter-wave (mmWave) bands, do not hold~\cite{6g_near_field_magazine,balanis2016antenna}. 

Here, we observe that the properties of a beam generated from an aperture are completely defined by the phase and, in some cases, amplitude distribution of the electric field at the aperture itself. Thus, the knowledge of the wavefront, i.e., the imaginary line that connects all the points of a wave with the same phase, is enough to completely characterize the beam~\cite{headland2018tutorial}. However, exploring the concept of wavefront engineering at THz frequencies to generate beams that have well-defined propagation characteristics in the near-field is a significant task, much more complicated than canonical phased array beamforming.
Aside from \emph{near-field beamfocusing} which utilizes a lens-like approach from the optical domain, other wavefronts and corresponding beams have also been explored, primarily for imaging and sensing purposes~\cite{durnin1988comparison,siviloglou2007observation}. The THz band however provides a Goldilocks zone for a novel implementation of such wavefronts in realizing efficient communication links as well.

In this paper, we present \emph{a vision for THz wavefront engineering to realize practical THz band communications} by utilizing wavefronts that can operate in the near field, mitigate the blockage problem, and guarantee multiple un-correlated channels matching perfect MIMO expectations, without the need of advanced signal processing algorithms or active relays. The present tutorial should boost further research in the growing area of combining electromagnetics and communications for physical layer design in 6G and beyond wireless systems.

\begin{figure*}[h!]
\begin{center}
\includegraphics[width=0.8\textwidth]{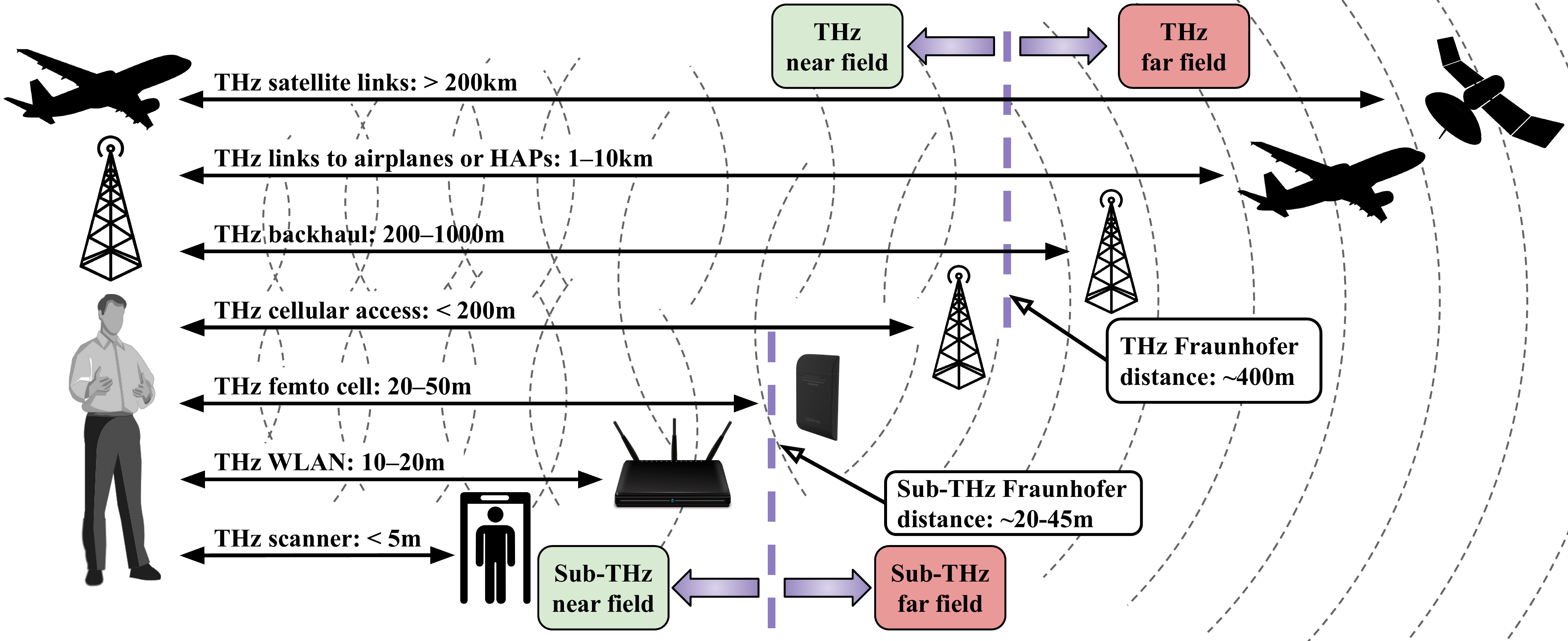}
\end{center}
\caption{Near-field and far-field 6G-grade sub-THz prospective use cases (300~GHz carrier and a 10-15~cm aperture at the larger unit or the base station) and beyond-6G true THz setups (1~THz carrier and a 25-cm aperture at the base station for increased coverage).}
\vspace{-2mm}
\label{fig:bandwidth}
\label{fig:near_field}
\end{figure*}

\section{The Inherent Problem of THz Communications}
\label{sec:problem}


Modern radio communications have been, for the most part, explored and developed under two major postulations: (i) the far-field assumption; and (ii) the narrow-band assumption. \emph{Prospective ultra-high data rate THz systems contradict both these~assumptions.}

\subsection{Near field vs. Far field}
The small wavelengths of THz signals result in small radiating elements. Thus, even moderately sized physical devices (up to a few cm) can pack many radiating elements, providing a high gain for overcoming large path losses. At the same time, this results in large electrical apertures (ratio of physical size to signal wavelength). The relative phase delays and signal strengths of the radiation from the elements of such apertures are configured such that their electromagnetic (EM) superposition creates the desired beam profile, such as concentrating the power in the desired spot/direction.

Under the far-field assumption, the resultant superimposed beam has a uniform phase across its cross-section, effectively like a plane wave. However, as the near-field region of an array increases exponentially with its physical size and linearly with the frequency, the far field Fraunhofer distance for practical THz communication links can easily be \emph{hundreds of meters away} from the transmitter~\cite{balanis2016antenna}. Hence, many prospective THz use cases will operate in the near field, as illustrated in Fig.~\ref{fig:near_field}.

\subsection{Wideband and MIMO issues}
The second fundamental assumption that does not always hold at THz frequencies is the narrow-band assumption. Following it, the bandwidth of the signal is considered negligibly small compared to the central frequency; the entire signal can be approximated as a single ``wave" at this design frequency. 

However, one of the key motivations in 
 exploring new spectrum is to achieve extremely high data rates and sensing resolution by harnessing tens or even hundreds of gigahertz (GHz) of the spectrum. Thus, the narrow-band assumption may not hold, and the same phase delay across the array elements no longer provides a uniform time delay across the signal bandwidth when moving away from a central frequency. This effect is also present in mmWave communications but becomes more profound in wider-band THz systems~\cite{akyildiz2022terahertz}.

Exploitation of MIMO techniques such as spatial multiplexing is one way to address the wideband-related difficulties without compromising on the desired capacity. 
However, the sub 6 GHz MIMO point-to-point spatial multiplexing methods that rely on rich multipath may not be viable in THz channels. Specifically, preliminary THz MIMO studies indicate that only a few dominant paths of communication exist in the THz channel, and are strongly correlated, thus making the THz channel strongly Rician, low rank, and sparse~\cite{han2021hybrid}.  
Further still, communication devices would still require an advanced THz MIMO architecture with powerful back-end processing for processing the independent streams accurately, naturally increasing the system complexity and costs even further~\cite{akyildiz2022terahertz}.

\subsection{Blockage of the THz signal}
Another inherent limitation of THz propagation is that THz signals are easily blocked by many objects in the environment, including building walls, furniture, and even parts of the human body. 
Today, the default approaches to mitigate blockage obstructions are: (i) applying IRSs to extend the coverage and engineer NLoS paths~\cite{di2019smart}; and (ii) exploiting THz multi-connectivity solutions~\cite{akyildiz2022terahertz}. 
Both approaches are feasible 
but come with an increased system design complexity.

\section{Wavefront Engineering\\to Conquer The THz Channel}
\label{sec:vision}
\begin{table*}
    \centering
    \setlength\tabcolsep{2pt}
       \caption{Comparison between different wavefront candidates for THz communications}
    \begin{tabular}{p{0.1\textwidth}p{0.15\textwidth}p{0.1\textwidth}p{0.1\textwidth}p{0.1\textwidth}p{0.1\textwidth}p{0.15\textwidth}}
        \toprule
        \multirow{1}{*}{Type} & \multirow{1}{*}{Phase profile} & \multicolumn{2}{c}{Beam profile} & \multirow{1}{*}{Self-healing} & \multirow{1}{*}{OAM} & \multirow{1}{*}{Self-accelerating} \\
        \cmidrule{3-4}   
        {} & {} & Near field & Far field & {}\\
        \midrule
       Beamforming & Planar & Undefined & Gaussian & \large{X} & \large{\checkmark} & \large{X}\\
       Beamfocusing & Quadratic & Point & Undefined & \large{X} & \large{X} & \large{X}\\
       Bessel beams & Conical & Bessel & Annular & \large{\checkmark} & \large{\checkmark} & \large{X}\\
       Airy beams & Exponential & Airy & Undefined & \large{\checkmark} & \large{\checkmark} & \large{\checkmark}\\
        \bottomrule
    \end{tabular}
    \label{tab:wavefront_comparison}
\end{table*}

Here, we introduce the vision of utilizing novel wavefronts to realize robust THz communications. The discussed THz wavefronts are illustrated in Fig.~\ref{fig:wavefronts_fig2} and their key features are summarized in Table~\ref{tab:wavefront_comparison}.

\subsection{Wavefront Generation}
\label{subsec:vision_generation}
The Huygens-Fresnel principle shows that the characteristics of a generated beam are completely defined by the phase and amplitude distribution of the electric field at the radiating aperture~\cite{headland2018tutorial}. Therefore, to generate any of the beams, all that is required is the corresponding wavefront which can be engineered by a specific phase profile. In addition to implementing this through an array, the same is also possible through custom-designed lenses, or in reflection through the utilization of reflectarrays and metasurfaces. A more in-depth discussion of these different design implementations is presented in Sec~\ref{sec:challenges}.

\subsection{Beamfocusing and beamforming}
\label{subsec:vision_beamforming}
Beamfocusing and beamforming are two canonical, interrelated wave types in the near field and far field, respectively. With \emph{beamfocusing}, a transmitting array is configured with a lens-like quadratic phase profile to focus the signal at a particular focal point, at which point the wavefront converges to a singularity~\cite{6g_near_field_magazine}. Beam intensity is maximum at this focal point, and the Abbe limit, which dictates the resolution capability of the array, gives the resultant beam spot size.

\emph{Beamforming} is the far-field extension of beamfocusing, such that the focal point is, effectively, moved to infinity -- the quadratic phase becomes a constant, uniform phase and the wavefront becomes planar, resulting in a beam that diverges in the far-field due to limited aperture size. Both beams have their limitations. Beamfocusing is focused at a given point, thus requiring perfect CSI and mobility tracking, while beamforming is valid only in the far field. Neither of these has any strong resilience to blockage.


\subsection{Bessel beams}
\label{subsec:bessel}
When discussing the candidate solutions for near-field THz systems, one of the solutions constitutes a beam intensity profile given by a zeroth-order Bessel function, giving these their name -- \emph{``Bessel beams"}~\cite{durnin1988comparison}. A radially symmetric linear phase creates this Beam profile, resulting in a conical wavefront.

\begin{figure}[!t]
\centering
	\includegraphics[width=0.9\linewidth]{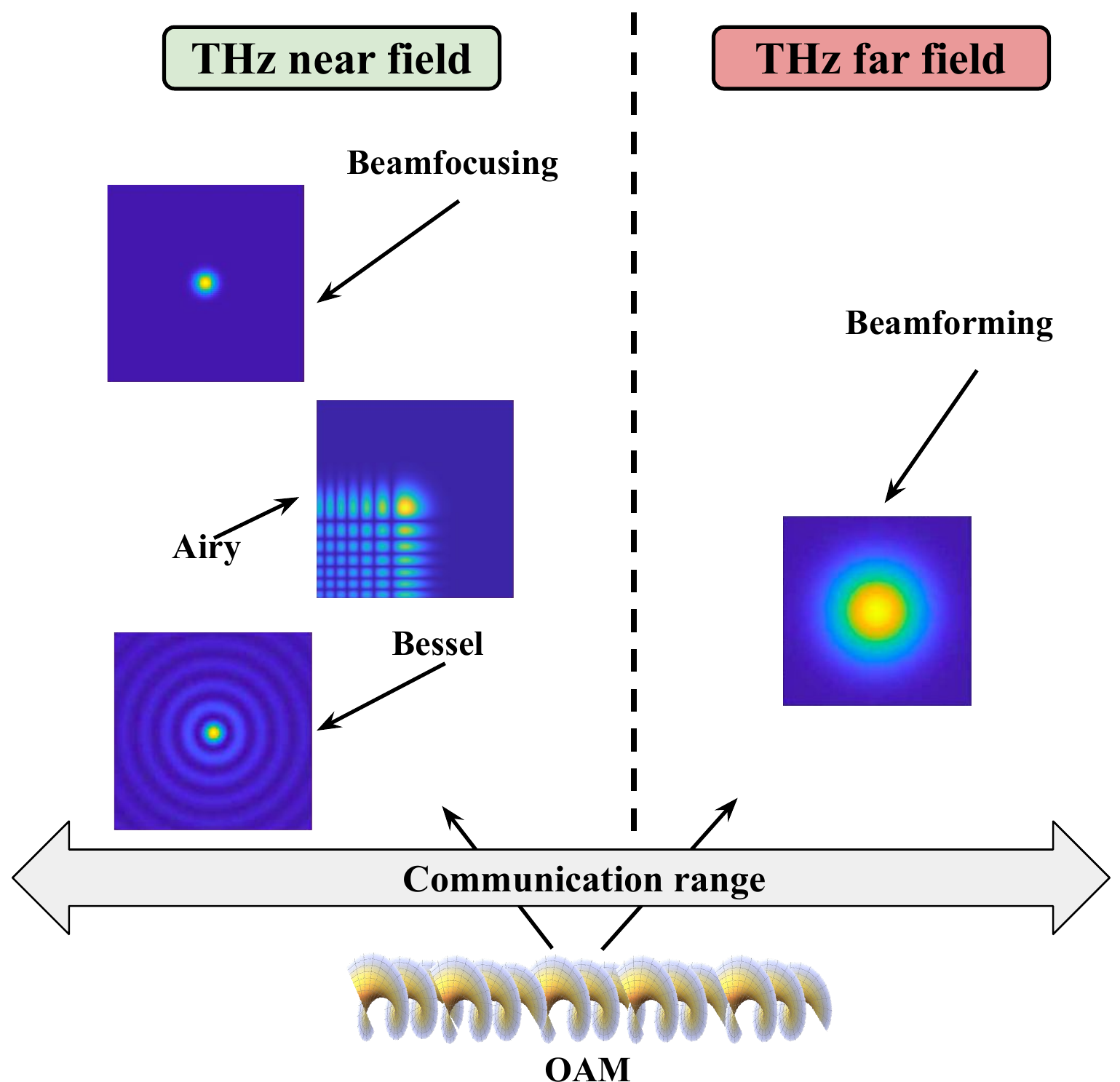}
	\vspace{2mm}
	\caption{Wavefront engineering for creating wireless THz links, in which the beam profiles can significantly deviate from the Gaussian.}
	\label{fig:wavefronts_fig2}
	\label{fig:system_model}
\vspace{-1\baselineskip}
\end{figure}

As illustrated in Fig.~\ref{fig:wavefronts_fig2}, the zero-order Bessel beam profile has a central bright spot along the cone axis, with multiple concentric rings around it, which are interference patterns created from the interference of plane waves from the opposite sides of this central axis. 
Bessel beams are exact solutions to Maxwell's equations with a constant beam intensity. Thus, as the wave propagates, the beam profile remains invariant (no spreading losses). The phase and total power within the central spot and all the concentric rings are the same, with much greater intensity in the central spot. However, for a finite aperture and power, the propagation distance over which a practical Bessel beam retains its properties is limited. 
Some of the major design considerations for Bessel beams are that, for a given aperture size, the angle of the conical wavefront directly affects the resultant diameter of the central spot, the number of rings, and the maximum propagation distance. Further, the electrical size of the aperture also factors into these considerations and all can be customized. For example, a particular design constraint is that Bessel beams generated from the same aperture but with a larger number of rings theoretically lead to better self-healing capabilities, while a smaller number of rings and a wider central bright spot leads to greater propagation distance~\cite{durnin1988comparison}.

Bessel beams propagate in the near field, a feature absent in both beamforming (not valid in the near field) and beamfocusing (does not propagate but converges at a focal point). In addition, as Bessel beams can be understood to be the interference pattern of plane waves traveling inwards on a cone, they are resilient to blockage; even if some of the waves are blocked by an obstruction, the remaining waves still reconstruct the interference pattern and help regenerate the Bessel beam after the obstruction; hence these are also referred to as \emph{self-healing beams}~\cite{li2017adaptive}. These properties make Bessel beams a promising wavefront candidate for THz communications.

\subsection{Curved and Airy beams}
\label{subsec:airy}
While Bessel beams can ``plow'' through an obstacle, there is another class of beams that can completely circumvent blockage by following a curved trajectory. When viewed in the transverse direction, the beams appear to have acceleration without any external energy -- giving them the name of \emph{self-accelerating beams}. Such beams have been recently demonstrated in EM applications~\cite{siviloglou2007observation}, where the beam profile can be described with the Airy function. The resultant Airy-like beam is a subset of self-accelerating beams, with the cross-sectional intensity illustrated in Fig.~\ref{fig:wavefronts_fig2}.

Using the concept of caustics from ray optics, an arbitrary curve can be generated from a large input aperture where the local required phase, and subsequent wavefront, are found using tangents of the curve extended to the aperture. A particular design consideration of these beams is that greater beam curvature requires both a steeper phase progression across the array aperture as well as a larger aperture size. In THz wireless communications, their ability to follow almost any curved, or convex, path makes them very promising to avoid signal blockage. In addition, accelerating beams also have the potential to change the angle of arrival at the receiver side, making them useful during beam alignment.

\subsection{OAM -- A New Feature}
\label{subsec:vision_OAM}
Certain wavefronts possess new properties which open the door to new communication strategies. The most widely researched and explored among these is orbital angular momentum (OAM), a property of light first demonstrated in~\cite{allen1992orbital}. A beam that has OAM manifests a spiral phase in the transverse direction, resulting in a helical wavefront with a zero-intensity vortex in the center, as shown in Fig.~\ref{fig:wavefronts_fig2}.

Beams with distinct OAM modes are always orthogonal to each other, and the resulting cross-correlation is guaranteed to be zero. Therefore, OAM mode multiplexing can be utilized to create simultaneous uncorrelated data streams~\cite{zhou2022utilizing}.  
Gaussian, Bessel, and Airy beams can all be customized to carry OAM by adding a spiral phase on top of the existing beam profile. 
Thus, in the near field, Bessel beams with OAM further augment the attractive features of Bessel beams~\cite{li2017adaptive}.

\subsection{Receiver Performance and Operation}
\label{subsec:vision_receiver}
Employing wavefront engineering at the transmitter and exploiting the properties of the discussed beams can truncate some of the more significant drawbacks of the THz channel, such as the near-field effect, blockage, and wideband-related difficulties. Ultimately, the objective is to provide greater signal power available at the receiver, thereby improving the SNR and performance metrics. 

Here, we note that during near-field operation, both the magnitude of the signal reaching the receiver and its coherent reception at the receiver becomes important. For example, with beamfocusing, if the receiver is not at the focal spot but slightly closer to the transmitter, the incident power is spread over an arc. Hence, the phase is non-uniform across the receiver cross-section. The receiver must then be able to accommodate the required delay and sum method for coherently adding such a non-uniform power profile~\cite{6g_near_field_magazine}.

With THz Bessel beams, the entire Bessel intensity pattern has the same phase, thereby making the receiver operation similar to that within far-field beamforming~\cite{headland2018tutorial}. With THz Airy beams, the receiver should be tuned to be coherent towards the incident angle defined by the curve of the Airy beam (similar to the far-field beamsteering direction of arrival).

When required to demultiplex multiple OAM modes, the receiver must employ a large enough aperture with a specific phase-based diffraction grating that exploits the principle of zero cross-correlation in OAM modes for ease of demultiplexing~\cite{petrov4mobile}. Nonetheless, the fundamental principles remain the same; more power received coherently allows for greater SNR and performance metrics.   

\section{THz Wavefronts: A Benchmark Study}
In this section, we present a numerical study illustrating how the described wavefronts and beams can quantitatively assist in mitigating the challenges of THz band communications. We consider a planar array of a side length of 25~cm 
operating at 1~THz and designed with a typical half-wavelength spacing of elements. The nominal gain of such an array is 70~dBi, which can help sustain local area, multi-GHz wide links. 

\subsection{Improved Radiation Gain}
\label{subsec:results_gain}

\begin{figure}
\centering
\vspace{+1mm}
	\includegraphics[width=0.9\linewidth]{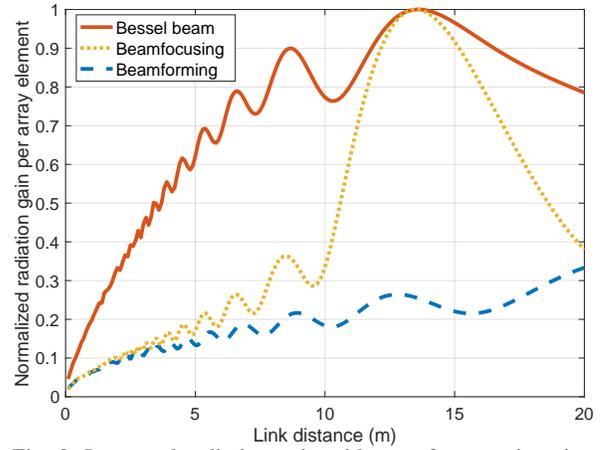}
	\caption{Improved radiation gain with wavefront engineering.}
	\label{fig:snr_results}
\vspace{-1.5\baselineskip}
\end{figure}

The effective normalized radiation gain is a key measure of how accurately an applied phase profile at the array constructively superimposes the waves from each of the array elements at a given evaluation point, and linearly affects the derived directivity of the array, with a maximum value of $1$. Both the received power and the SNR at the receiver are linearly proportional to the effective normalized radiation gain. 

In Fig.~\ref{fig:snr_results}, we present this normalized radiation gain under the application of Bessel beams, beamfocusing, and beamforming, all designed to steer in the broadside direction of the array. The peculiar beam profile of Airy beams cannot be grouped with the other beam shapes, since the Airy beam follows a curved trajectory. Nonetheless, for the path specified during wave generation, the Airy beam will have an intensity maximum along that curve. The diameter of the central spot of the Bessel is set to 20~mm, which, given the aperture size, results in a maximum propagation distance of 23~m. 

It is observed from Fig.~\ref{fig:snr_results} that under beamforming the normalized gain is notably crippled at distances of up to 20~m, which are typical ranges for prospective indoor THz WLANs. In contrast, the near field validity of Bessel beams results in a significantly improved radiation gain for the same settings. Arbitrarily, the focal point of beamfocusing is set to the distance where the normalized gain with the Bessel beam is maximized (at 13.5~m). The corresponding curves intersect at that point. However, even small deviations from this focal point (i.e., due to mobility) result in the efficiency of beamfocusing reducing rapidly and much faster than the Bessel curve. Hence, in the absence of a perfectly updating CSI (impractical in many setups), sustaining acceptable performance for mobile THz systems with beamfocusing is unlikely. The improvement in beamforming will be slightly faster for smaller apertures or lower frequencies since, in both cases, the electrical size of the aperture decreases. However, these changes simultaneously lead to a reduced aperture gain that challenges coverage and capacity. The qualitative system behavior remains the same: \emph{THz Bessel beams are more efficient for shorter distances (up to several tens of meters), while THz beamforming is more efficient for longer links closer to the far-field region.}

\subsection{Blockage Mitigation}
\label{subsec:results_blockage}

\begin{figure}
\centering
\vspace{+1mm}
	\includegraphics[width=0.9\linewidth]{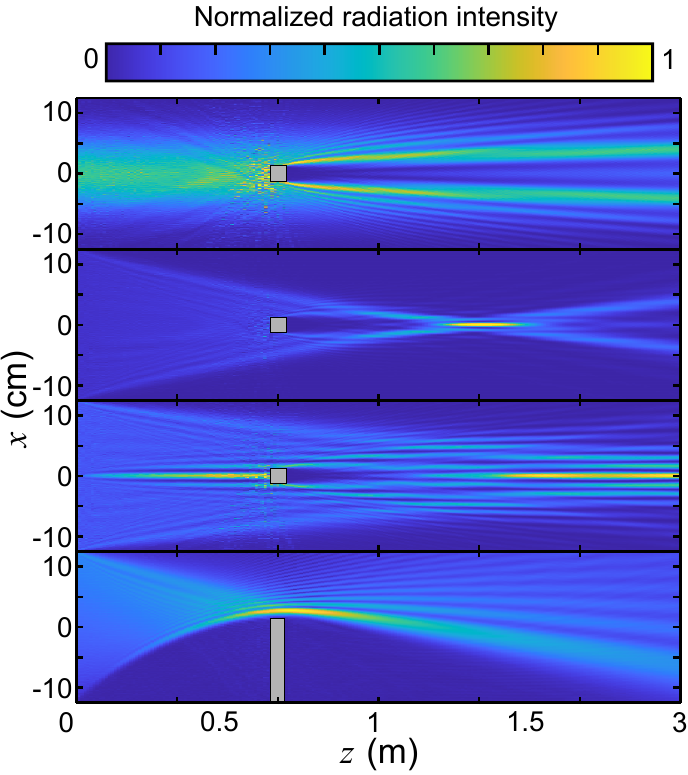}
	\caption{Blockage mitigation with different wavefronts. From top to bottom: beamforming, beamfocusing, Bessel, Airy-like curved beam. Full wave finite-element-method (FEM) with COMSOL Multiphysics is utilized.} 
	\label{fig:blockage}
\vspace{-1.5\baselineskip}
\end{figure}

The potential of wavefront engineering in mitigating the impact of obstacles to THz signal propagation is illustrated in Fig.~\ref{fig:blockage}. The wave is steered in the broadside and encounters an obstacle that is 10\% the size of the aperture, except in the case of the Airy-like beam which curves around a much larger, indefinitely extending obstacle. This also highlights additional possibilities in blockage mitigation with THz Airy beams.

With beamforming, the radiation intensity behind the obstacle is significantly diluted and the wave is blocked. However, other cases show certain resilience to blockage. With beamfocusing, the radiation from the portion of the aperture not blocked by the obstacle can still converge to focus at the desired focal point. However, this is not technically ``self-healing'', as the focal spot had not yet been formed; were the obstacle closer to the focal point, the blockage would be severe. In contrast, the self-healing nature of Bessel beams \emph{reforms the beam} beyond the obstacle and the radiation intensity is almost unaffected. The performance of the Airy-like beam also remains impervious to the blockage since the beam can be designed to curve around this obstruction.

\subsection{Increased System Capacity}
\label{subsec:results_multiplex}

\begin{figure}[!h]
\centering
\vspace{+1mm}
	\includegraphics[width=0.9\linewidth]{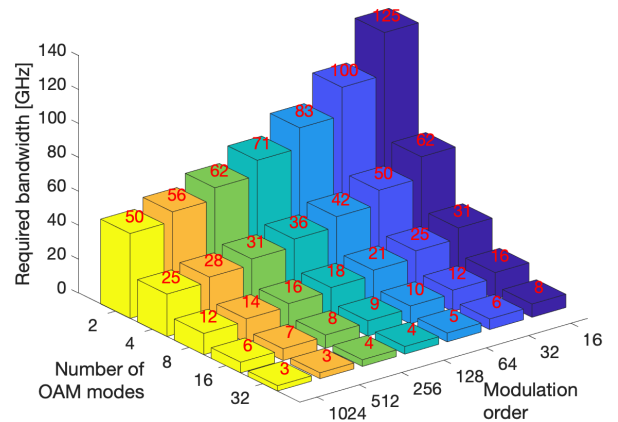}
	\vspace{0.3\baselineskip}
	\caption{Required bandwidth for a 1~Tbps link capacity with OAM-Multiplexing enabled.}
	\label{fig:OAM_spatial_reuse}
\end{figure}

The exploitation of wavefronts that carry OAM, such as higher-order Bessel beams in the near field, allows the creation of perfect parallel orthogonal channels. 
In Fig.~\ref{fig:OAM_spatial_reuse}, we show the required bandwidth to achieve a point-to-point 1~Tbps wireless link for different OAM configurations and various orders of quadrature amplitude modulation (QAM), assuming no distortions on the spiral phase for illustrative purposes. 

It is observed from Fig.~\ref{fig:OAM_spatial_reuse} that the demand on the bandwidth can be significantly reduced when both OAM and higher-order modulations are used. For example, utilizing 32 beams, each with a distinct OAM mode, the required bandwidth is less than 10~GHz, even with 16-QAM. This can be further reduced down to 3~GHz with a more sophisticated 1024-QAM modulation. 
The latter could not only facilitate the spectrum utilization and the co-existence of prospective THz links with other services (i.e., the Earth exploration satellites) but also allows to avoid certain wideband-specific challenges discussed earlier.

\section{Research Challenges and Opportunities}
\label{sec:challenges}

The efficient harnessing of novel wavefronts and their attractive properties within widespread 6G and beyond scenarios is a wide research area with many open challenges:

\subsubsection{Near-field THz channel}
The work should start with a better understanding of the near-field THz channel itself. 
Solving Maxwell's equations numerically for all the possible THz channel configurations is computationally extensive in many setups. Hence, there is a need to develop THz-specific channel models or approximations for near-field propagation in target setups, such as indoor and outdoor wireless access. These could not only account for the equivalent path loss in the near field but also scattering and reflection effects from possible complex propagation environments in the near field.

\subsubsection{Near-field THz sensing}
Adding sensing capabilities to wireless access systems is today considered one of the major innovations toward 6G. Nonetheless, the peculiarities of near-field propagation demand updating the relevant metrics and models. In addition, the discussed wavefronts (and other counterparts) have been previously utilized in sensing and imaging applications in the optical domain. When it comes to THz imaging and sensing, as well as the promising research topic of joint communications and sensing (JCS), the 
applicability and impact of wavefront engineering must be explored with the specifics of the underlying beam profile taken into account. Here, an interesting research problem is the fact that one of the important metrics in sensing, i.e., the radar cross-section (RCS) is, strictly speaking, defined in a far-field sense. While recent findings on near-field RCS utilizing the spherical wave model are summarized in~\cite{dogaru2021near}, evaluating the effectiveness of the developed near-field model for the other wavefronts is a pressing task. Further, in the advent that the model is not valid ubiquitously, developing a general solution that captures the significant complexity of these beams for near-field RCS could be a promising research avenue.

\subsubsection{Wavefront Distortions in Ultra-broadband THz Links}
One of the fundamental expectations in practical THz systems is enabling \emph{broadband} communications. Wavefront engineering is a narrow-band technique: a targeted feature of a wavefront does not extend uniformly across a large bandwidth, as the same phase delay does not generate the same time delay that truly governs wave propagation, superposition, and final beam shape. 
True time delay lines can be utilized to counter this problem through a frequency-dependent phase profile that makes the time delay uniform. However, the required architecture is more complex~\cite{headland2018tutorial}.

Here, OAM multiplexing can help boost the system capacity, matching optimal (uncorrelated) MIMO. In addition, OAM multiplexing/demultiplexing is less demanding from a signal processing perspective compared to THz MIMO, since, as mentioned in Sec.~\ref{subsec:vision_receiver}, the receiver only needs the correct phase profile across its aperture. Nonetheless, if the signal's bandwidth within the OAM mode is large enough, the generated OAM modes are no longer pure and can lead to crosstalk or interference between multiplexed OAM streams. Therefore, further advances are needed to properly characterize these new performance-complexity trade-offs, such as by characterizing and quantifying this spillover effect. 


\subsubsection{Mobile THz systems}
Mobility is one of the principal challenges when it comes to operating in practical, everyday scenarios. Both Bessel and Airy beams propagate in the near field, and thus the application of beamsteering techniques from the far field is also valid here. However, it has been shown that OAM-carrying beams, when steered, have a spillover effect that results in crosstalk (interference) with neighboring OAM modes. Additionally, to apply both steering and wave-generating phase profiles, the phase shifters need to be significantly complex; 
4-bit (16-level) phase shifters are usually sufficient for almost all applications~\cite{headland2018tutorial}, specifying the phase resolution threshold. On the positive side, the blockage mitigation capabilities may relax certain mobility-related requirements such as the degree of THz multi-connectivity~\cite{akyildiz2022terahertz}.

\subsubsection{Interference Modeling}
Most, if not all, studies on interference modeling for mmWave and THz communications assume plane waves~\cite{kth_interference_survey}. 
Thus, the results and observations from prior studies on interference in THz networks must be revisited to account for the wavefront specific properties.
These findings would be of paramount importance for efficient frequency and space reuse, medium access control (MAC), and resource allocation in prospective multi-user THz networks.

\subsubsection{Mechanisms  to Generate Wavefronts}
The generation and detection of custom wavefronts can be achieved with a lower complexity of the digital backends preceding the array apertures. Particularly, holographic beamforming architectures may significantly reduce the power consumption and cost in developing large-scale arrays~\cite{deng2021reconfigurable}. 
Moreover, the extreme scalability of IRSs can constitute large intelligent surfaces (LISs), which can extend the applicability of these wavefronts to large-scale cellular deployments.

This setup, however, depends on the accuracy of the phase-shifting elements (as discussed above) and the spatial resolution of the array~\cite{di2019smart}. Utilizing a metasurface approach with sub-wavelength radiating elements could increase the spatial resolution~\cite{di2019smart}. Nonetheless, the radiation response of such metasurfaces is dependent on the mutual coupling between these densely packed elements; accurately applying the phase profile for a desired wavefront is~challenging.

Clearly, for dynamic wavefront customization, the utilization of lenses and non-reconfigurable dielectrics will prove insufficient. However, the significant recent advances in reconfigurable arrays for THz communications indicate that dynamic and advanced wavefront engineering will be possible in the 6G and beyond landscape for improved mobility.

\subsubsection{Energy Efficiency}
The energy cost of THz communications is still a significant challenge, primarily due to the lack of efficient THz radiation generation in comparison to power generation at lower frequencies. In addition to significant advances in device technology~\cite{akyildiz2022terahertz}, wavefront engineering can also be a potential solution since the beam energy can be better focused towards the intended directions. The substantial energy challenge of blockages can be addressed by leveraging the self-healing and self-accelerating properties of Bessel and Airy beams. Still, comparative research is required to identify which of the beams to use to address this problem.

\section{Conclusion}
\label{sec:conclusions}

With continuous growth in user demands, the THz band is envisioned to be explored for both high-rate data exchange and high-precision sensing in next-generation wireless systems. However, the dogmas of conventional RF systems from the microwave and mmWave bands cannot hold infinitely for higher frequencies. Particularly, the THz physical layer must have a symbiosis of communication and wave theory. Here, the design of novel custom THz wavefronts discussed in this article is one of the key enablers for efficient practical THz communication systems that will constitute an inherent part of the 6G and beyond-6G landscapes.



\ifCLASSOPTIONcaptionsoff
  \newpage
\fi


\balance
\section*{Authors' Biographies}

\textbf{Arjun Singh} (singha8@sunypoly.edu) is an Assistant Professor in the Department of Engineering at the State University of New York Polytechnic Institute, Utica, NY, USA. 
His research interests include realizing THz band wireless communications for 6G and beyond.

\textbf{Vitaly Petrov} (v.petrov@northeastern.edu) is an Associate Research Scientist, Northeastern University, Boston, MA, USA. 
His research interests include THz band communications for 6G and beyond.

\textbf{Hichem Guerboukha} (hichem\_guerboukha@brown.edu) is a Postdoctoral Research Fellow with the School of Engineering, Brown University, Providence, RI, USA. 
His research interests include THz instrumentation and waveguides, THz computational imaging, and THz communications.

\textbf{Innem V.A.K. Reddy} (innemven@buffalo.edu) is a dual-Ph.D. graduate from the University at Buffalo, SUNY, NY, USA, and King Abdullah University of Science and Technology, Saudi Arabia. His current research interests are in the field of micro-optics, beam shaping, 3D nanofabrication, optical tweezers, plasmonics, and THz communications.

\textbf{Edward W. Knightly} (knightly@rice.edu) is the Sheafor-Lindsay Professor of Electrical and Computer Engineering at Rice University, Houston, TX, USA. 
His research interests are in experimental wireless networking and wireless network security.

\textbf{Daniel M. Mittleman} (daniel\_mittleman@brown.edu) is a Professor in the School of Engineering at Brown University. 
His research interests involve the science and technology of THz radiation. 

\textbf{Josep Miquel Jornet} (jmjornet@northeastern.edu) is an Associate Professor in the Department of Electrical and Computer Engineering at Northeastern University. 
His research interests are in THz communications and wireless nano-bio-communication networks. 
\end{document}